\documentclass[twocolumn,pra,showpacs,amsmath,amssymb]{revtex4}
\usepackage{amsmath,amsfonts,amssymb,graphics,graphicx,epsfig,color,times,bbm,natbib}

\begin{document}

\title{Concavity of the quantum body for any given dimension}
\author{K\'aroly F.~P\'al}
\email{kfpal@atomki.hu}
\author{Tam\'as V\'ertesi}
\email{tvertesi@dtp.atomki.hu}
\affiliation{Institute of Nuclear Research of the Hungarian Academy of Sciences\\
H-4001 Debrecen, P.O.~Box 51, Hungary}

\def\CC{\mathbb{C}}
\def\RR{\mathbb{R}}
\def\one{\leavevmode\hbox{\small1\normalsize\kern-.33em1}}
\newcommand*{\tr}{\mathsf{Tr}}
\newcommand{\diag}{\mathop{\mathrm{diag}}}
\newcommand{\ket}[1]{|#1\rangle}
\newcommand{\bra}[1]{\langle#1|}
\newtheorem{theorem}{Theorem}
\newtheorem{lemma}{Lemma}

\date{\today}

\begin{abstract}
Let us consider the set of all joint probabilities generated by local binary measurements on two separated quantum systems of a given local dimension $d$.
We address the question of whether the shape of this quantum body is convex or not. We construct a point in the space of joint probabilities, which is
on the convex hull of the local polytope, but still cannot be attained by measuring $d$-dimensional quantum systems, if the number of measurement
settings is large enough. From this it follows that this body is not convex. We also show that for finite $d$ the quantum
body with POVM allowed may contain points that can not be achieved with only projective measurements.
\end{abstract}

\maketitle

Quantum correlations are points in the space of correlations which are achievable in quantum phyics by performing local measurements on separate quantum systems.
In contrast, classical correlations can be achieved by local strategies using shared randomness. For a given number of measurement inputs and outputs 
the set of classical correlations forms a convex polytope \cite{Froissart}. However, we have learned from the theorem of John Bell that there exist quantum correlations
that lie outside this polytope \cite{Bell}. 
Thus the set of quantum correlations (which we refer to as the quantum body) is strictly larger than the set of classical correlations.

Let us first consider the quantum body consisting of two spacelike separated parties, each having a choice of performing two measurements with two outputs. If we keep only joint correlations in the set (excluding marginal terms), we obtain the simplest nontrivial quantum body. 
The boundary of this set has already been described by Tsirelson \cite{Tsirelson}, notably, deriving the maximal quantum violation for the
Clauser-Horne-Shimony-Holt (CHSH) inequality \cite{CHSH}. 
Subsequent works \cite{corr22} characterized the boundary of this quantum domain in different but essentially equivalent ways. Recently, the structure
of this body has been the subject of analytical study in Ref.~\cite{Pit08} deriving quadratic inequalities. However, the fact that in this quantum body
marginal terms are not included, the inequalities derived can give only partial information on the full probability distribution, i.e., on the shape of
the whole quantum body for two parties with two inputs and two outputs. 

Beyond this scenario, Navascu\'{e}s et al.~\cite{NPA07} devised a sophisticated method based on a hierarchy of semidefinite relaxations. This is completely
general, in that it can be applied to any number of parties, performing measurements with any number of  inputs and outputs. However, the method in its present form works efficiently for
the case when no constraints are imposed on the dimension of the system. In absence of such a powerful program, the shape of the quantum body for a
fixed dimension is not well understood. It is not even known whether it is convex or concave. Without the restriction for the dimensionality of the
quantum systems the quantum body is proven to be convex~\cite{Pit89,WW}.
It is also known, that the size of the quantum body may grow with $d$ for two parties \cite{dimwitness,VP09,higherdim} and for three parties as well \cite{Perez}.

In the present paper we wish to further advance the study on the shape of the quantum body corresponding to a fixed Hilbert space dimension $d$
of bipartite systems. We address the problem recently raised by Navascu\'{e}s et al.~\cite{NPA08}: Is the shape of the quantum body convex for a
restricted dimension $d$? Our main result shows that even by four two-outcome measurement settings per party the corresponding quantum body for
a pair of two-dimensional quantum systems (qubits) is concave. This result holds for the most general POVM measurements and for projective
measurements as well and can be generalized beyond qubits to any dimension $d$.

\emph{Preliminaries.}
Let Alice and Bob have two components of a compound physical system. Let Alice and Bob choose one of a set
of $m_A$ and $m_B$ two-outcome measurements, respectively, and let them perform the measurement
chosen on their respective subsystems. Let us denote the outcome of Alice's measurement $i$
and Bob's measurement $j$ by $A_i=\pm 1$ ($i=1,\dots,m_A$) and $B_j=\pm 1$ ($j=1,\dots,m_B$),
respectively. Let us denote the vector having components $\langle A_i\rangle$, $\langle B_j\rangle$
and $\langle A_iB_j\rangle$ for all $i$ and $j$ by $\bf x\in\RR^{m_A+m_B+m_Am_B}$, where
$\langle\cdot\rangle$ denotes the expected value. Vector $\bf x$ may be measured
by repeating the procedure above on many copies of the system, making sure that each pair
of measurements $(i,j)$ is chosen to be performed a sufficient number of times to get a satisfactory statistics.
The actual vector one gets will depend on the physical system and on the measurement settings
the parties are allowed to choose from. We note that $\langle A_i\rangle$ and
$\langle B_j\rangle$ is only defined sensibly if the
probability of getting a measurement outcome by one party does not depend on which measurement the
other party has chosen. This is the requirement of no-signaling, which is true in both
classical and quantum physics, and believed to be true in Nature.

As we have mentioned above, the set of vectors one may get when making
measurements on systems obeying classical physics, or any locally realistic model, is a polytope \cite{Froissart,Pit89}.
The vertices of the polytope correspond to the deterministic situations, when each $A_i$ and
$B_j$ has a definite value every time it is measured, and the polytope itself is the
convex hull of these points. A Bell inequality of the form $\bf M\cdot\bf x\le K$ define an
$(m_A+m_B+m_Am_B-1)$-dimensional hyperplane touching the polytope, such that the
polytope is on that side of the hyperplane, which satisfies the inequality. Tight Bell inequalities
are the ones that define the hyperplanes of the facets of the polytope. While all vectors $\bf x$ allowed
classically may be reproduced by measurements on quantum systems, the opposite is not true.
In quantum mechanics Bell inequalities may be violated, therefore the set of the vectors $\bf x$
allowed is larger. It is not a polytope, but it is still a convex set \cite{Pit89,WW}.

In quantum mechanics the components of $\bf x$ may be calculated as
$\langle A_i\rangle={\rm tr}(\rho\hat A_i\otimes\hat I_B)$,
$\langle B_j\rangle={\rm tr}(\rho\hat I_A\otimes\hat B_j)$, and
$\langle A_iB_j\rangle={\rm tr}(\rho\hat A_i\otimes\hat B_j)$. Here $\hat A_i\in{\cal H}_A$
($\hat B_j\in{\cal H}_B$) is the observable corresponding to Alice's (Bob's) measurement $i$ ($j$),
${\cal H}_A$ (${\cal H}_B$) is the Hilbert space associated to the subsystem of Alice (Bob),
$\hat I_A\in{\cal H}_A$ and $\hat I_B\in{\cal H}_B$ are the unity operators of the respective
Hilbert spaces, and $\rho\in{\cal H}_A\otimes{\cal H}_B$ is the density operator of the physical
system. Operators $\hat A_i=\hat I_A-2\hat P^A_i$ and $\hat B_j=\hat I_B-2\hat P^B_j$ have eigenvalues
$\pm 1$, while $\hat P^A_i\in{\cal H}_A$ and $\hat P^B_j\in{\cal H}_B$ are projection operators,
whose expected values give the probability of getting outcome $-1$ for the corresponding measurements.
If we do not confine ourselves to projective measurements, but we allow the more general POVM
measurements, then $\hat P^A_i$ ($\hat P^B_j$) will be the POVM element associated to outcome $-1$ of
the Alice's (Bob's) measurement $i$ ($j$), which is not necessarily a projector, but any positive operator
with eigenvalues between $0$ and $1$. The relation with $\hat A_i$ ($\hat B_j$) remains the same as above.

\emph{Method.}
Let us arrange $X_{00}\equiv 1$, and the components of $\bf x$ into a matrix of $m_A+1$ rows and $m_B+1$
columns as $X_{i0}=\langle A_i\rangle$, $X_{0j}=\langle B_j\rangle$ and
$X_{ij}=\langle A_iB_j\rangle$. The quantum mechanical expression for these matrix
elements is $X_{ij}={\rm tr}(\rho\hat A_i\otimes\hat B_j)$, with the definitions
$\hat A_0\equiv\hat I_A$ and $\hat B_0\equiv\hat I_B$, here we allowed indices $i$ and $j$ to take value $0$.
Let us restrict the dimensionality of the component Hilbert spaces ${\cal H}_A$ and ${\cal H}_B$ to two.
In two dimensions all Hermitian operators can be written
as a real linear combination of the three Pauli operators and the unity operator, therefore if $m_A>3$,
there must exist an $\hat A_i$ operator which can be written as a linear combination of the other
$\hat A_k$ operators and the unity operator with real coefficients. Then the row of the $X_{ij}$ matrix
depending on that operator can also be written as the linear combination of the rows depending on
the other $\hat A_k$ operators and the zeroth row, with the same coefficients. We can conclude that
correlations described by $X_{ij}$ having more than four linearly independent rows can not be
reproduced with measurements taken on a pair of qubits. We may repeat the argument for the columns, too.
When we go beyond qubits, but still restrict ourselves to finite dimensional Hilbert spaces, we can draw
a similar conclusion: $X_{ij}$ having more than $d^2$ linearly independent
rows or columns can not be reproduced by measurements performed on systems with no more
than $d$-dimensional component Hilbert spaces. This is because any $d$-dimensional Hermitian matrix
can be characterized by $d$ real (diagonal elements) and $d(d-1)/2$ complex (nondiagonal elements) numbers,
altogether $d^2$ real numbers, therefore, no more than $d^2$ of them may be linearly independent with real
coefficients. The conclusion holds for both projective and POVM measurements, as only the hermiticity of
the operators has been used in the argument. The inclusion of POVM is important, because as we will
show, if the dimensionality of the quantum system is restricted, the quantum body can be larger if
we allow POVM. 

\emph{Main result.}
Now let us consider the set of $\bf x$ vectors achievable with
measurements on quantum systems of at most $d$-dimensional component Hilbert spaces.
The set is not convex, if there exist points in the vector space that belong to the set,
but some point on the convex hull of these points does not. The latter can be proven by showing that
the matrix $X_{ij}$ corresponding to that point has more than $d^2$ linearly independent columns or rows.
We will prove below that if $m_A=m_B=m\ge d^2$ and even, the set will not even
contain some vector on the convex hull of points corresponding to deterministic cases, that is an
element of the local polytope. The deterministic cases can obviously be reproduced with any
physical systems by using degenerate measurements, measurements with definite outcomes independent of the
physical system.

We note that to express the classical or quantum limits on results of correlation experiments very often
not the vectors $\bf x$, but the vectors $\bf p$ are used, whose components are $p_{A_i}$, $p_{B_j}$ and $p_{A_iB_j}$,
which are the probabilities of getting outcome $-1$ for Alice's $i$th, for Bob's $j$th, and
for both experiments, respectively. The two approaches are equivalent \cite{AII}.

Let $m_A=m_B=m$ be even, and let us take all $m!/(m/2)!^2$ deterministic cases with the $A_i$ being +1 the
same number of times as it is -1, and $B_i=-A_i$. Let us call the corresponding vectors
and matrix elements ${\bf x}^{(\sigma)}$ and $X_{ij} ^{(\sigma)}$ ($\sigma=1,\dots,m!/(m/2)!^2$), respectively.
Let ${\bf x}^{(+)}$ and ${\bf x}^{(-)}$ be defined by $A_i^{(+)}=B_i^{(+)}=+1$ and $A_i^{(-)}=B_i^{(-)}=-1$,
respectively. Let us take the following point on the convex hull of these vectors:
\begin{equation}
{\bf x}^o=\frac{m-1}{m}\frac{(m/2)!^2}{m!}\sum_\sigma{\bf x}^{(\sigma)}+\frac{1}{2m}\left ({\bf x}^{(+)}+{\bf x}^{(-)}
\right ).
\label{xcombin}
\end{equation}
For each deterministic strategy considered in the sum above, there is another one with the same weight
with all measurement outcomes having the opposite sign, therefore $X^o_{i0}=X^o_{0j}=0$ ($i,j=1,\dots,m$).
As $X_{ii} ^{(\sigma)}=A_{i} ^{(\sigma)}B_{i} ^{(\sigma)}=-1$, and $X_{ii} ^{(+)}=X_{ii} ^{(-)}=+1$, it follows that
$X_{ii}^o=(2/m)-1$. It is easy to see, that if $i\neq j$, the value of
$X_{ij} ^{(\sigma)}=A_{i} ^{(\sigma)}B_{j} ^{(\sigma)}=-A_{i} ^{(\sigma)}A_{j} ^{(\sigma)}$ is $+1$ for
$2[(m-2)!]/(m/2-1)!^2$ cases, and $-1$ for the rest of them, and it is obvious that $X_{ij} ^{(+)}=X_{ij} ^{(-)}=+1$.
From these and from Eq.~(\ref{xcombin}) it follows that the nondiagonal matrix elements with indices larger
than zero are $X_{ij}^o=2/m$. The matrix has a nonzero determinant, all $m+1$ rows and columns are linearly
independent, therefore, if $d^2\le m$, ${\bf x}^o$ can not be reproduced by measurements performed on
quantum systems with $d$-dimensional component Hilbert spaces.

\emph{Explicit Bell polynomial.}
Now we will show that all vectors ${\bf x}^o$, ${\bf x}^{(\sigma)}$, ${\bf x}^{(+)}$ and ${\bf x}^{(-)}$ considered
above belong to a set that maximizes a Bell inequality, which can not be violated in quantum mechanics,
so they are on the surface of both the classical polytope and the quantum set. As the quantum set has a multidimensional
intersection with the polytope, it follows that its surface can not be round everywhere. This fact has been also reported
recently in the work of Linden et al. \cite{Linden} in the context of distributed computing. The intersection has
a lower dimensionality than a facet, so the Bell inequality is not a tight one.
It is a correlation type inequality, that is the factors multiplying $\langle A_i\rangle$ and $\langle B_j\rangle$
are zero. The Bell polynomial is
\begin{equation}
{\cal B}=\sum_{i=1}^m\sum_{j=1}^m M_{ij}\langle A_iB_j\rangle\equiv\sum_{i=1}^m\sum_{j=1}^m
\left (1-\frac{m}{2}\delta_{ij}\right )\langle A_iB_j\rangle,
\label{bellexpr}
\end{equation}
where $\delta_{ij}$ is the Kronecker delta. To get the maximum value of this expression it is enough to
consider pure states and projective measurements. It is proven in \cite{AGT06} that for any observables
$\hat A$ and $\hat B$ in Alice's and Bob's component spaces, respectively, and state $\psi$ there exist
Euclidean vector $\vec a$ independent of $\hat B$ and vector $\vec b$ independent of $\hat A$ such that
$\langle AB\rangle=\langle\psi|\hat A\otimes\hat  B|\psi\rangle=\vec a\cdot\vec b$. Therefore,
we may replace $\langle A_iB_j\rangle$ with $\vec a_i\cdot\vec b_j$ in Eq.~(\ref{bellexpr}), and maximize
that expression. The vectors $\vec a_i$ have to be chosen such that they are parallel with the vectors they
are multiplied with. Then we get:
\begin{align}
\vec a_i&={\frac{1}{l_i}}\left (\sum_{j=1}^m\vec b_j-\frac{m}{2}\vec b_i\right ),\label{lconc1}\\
{\cal B}&=\sum_{i=1}^ml_i=\sum_{i=1}^m\left|\sum_{j=1}^m\vec b_j-\frac{m}{2}\vec b_i\right|=\label{lconc2}\\
&=\sum_{i=1}^m\sqrt{\frac{m^4}{4}+\sum_{j=1}^m\vec b_j\cdot\sum_{k=1}^m\vec b_k-m\vec b_i\cdot\sum_{k=1}^m\vec b_k}.\label{lconc3}
\end{align}
We will show that we get the maximum value for $\cal B$ if
\begin{equation}
(\vec b_i-\vec b_j)\cdot\sum_{k=1}^m\vec b_k=0\label{veccond}
\end{equation}
is true for any $i$ and $j$. Then one can see from Eqs.~(\ref{lconc2},\ref{lconc3}) that $l_i=m/2$,
and ${\cal B}=m^2/2$. This agrees with the upper limit this Bell expression may take with quantum measurements,
as it can be shown analytically making use of semidefinite programming technique. The actual proof, following Wehner's work \cite{Wehner} is deferred to Appendix~\ref{SDP}.

From Ref.~\cite{VP09} it follows, that if $m_A=m_B$, and the maximum value of the Bell expression can be achieved with all
$\vec b_j$ are linearly independent, than this solution can not be unique. The present case is an example for this situation.
Equation~(\ref{veccond}) has an infinite number of solutions, with $\vec b_j$ spanning spaces of any dimensionality up to $m$.
An obvious one-dimensional solution is when all $\vec b_j$ are chosen to be the same unit vector $\vec b$.
Then from  Eq.~(\ref{lconc1}) and $l_i=m/2$ it follows that $\vec a_i=\vec b$. This arrangement corresponds to the classical
deterministic strategies of having all measurement outcomes either $+1$ or $-1$ every time (correlation vectors
${\bf x}^{(+)}$ and ${\bf x}^{(-)}$).
If $m$ is even, there are further one dimensional solutions, with half the $\vec b_j$ pointing to one direction and the other half
pointing to the opposite direction. Such a solution corresponds to a deterministic strategy in which
Bob has as many measurements with a definite outcome of $+1$ as ones with an outcome of $-1$,
and Alice gets the outcome $A_i=-B_i$ for each $i$ (correlation vectors ${\bf x}^{(\sigma)}$).
From the existence of classical deterministic strategies giving the quantum limit
for the Bell expression it follows that the Bell inequality can not be violated.

There is an infinite number of solutions of Eq.~(\ref{veccond}) with $\vec b_j$ spanning the maximum of $m$ dimensions.
An arrangement with all $\vec b_j$ are orthogonal to each other is one of them.
Then $\vec a_i$ are also orthogonal to each other, and $\langle A_iB_j\rangle=\vec a_i\cdot\vec b_j=(2/m)-\delta_{ij}$,
(see Eq.~(\ref{lconc1})). According to Tsirelson's construction these values can be realized as quantum expectation values of
$\pm 1$ valued observables with a maximally entangled state of a system of $2^{m/2}$ dimensional component Hilbert spaces \cite{Tsirelson}.
With this state the expectation values $\langle A_i\rangle=0$ and $\langle B_j\rangle=0$. By choosing the unit vectors corresponding to
the unity operators in the component Hilbert spaces orthogonal to the space spanned by $\vec b_j$, we can get all components of the
correlation vector as dot products. This correlation vector is nothing else than ${\bf x}^o$, which we have chosen to construct on the
convex hull of the set of classical deterministic cases  ${\bf x}^{(\sigma)}$, ${\bf x}^{(+)}$ and ${\bf x}^{(-)}$ according to Eq.~(\ref{xcombin})
as an example that can not be achieved with quantum systems of component spaces of $d\leq\sqrt m$ dimensions.
Clearly, we could have chosen an infinite number of other vectors with the required property.

\begin{figure}
\vspace{-1cm}
\includegraphics[width=\columnwidth]{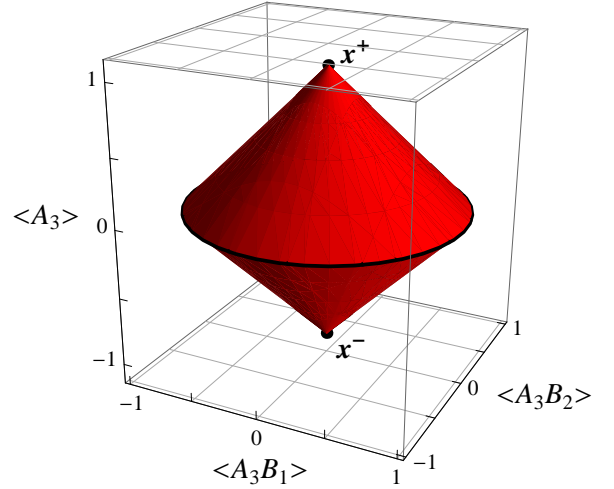} \caption{(Color online)
Quantum region in the three-dimensional section spanned by the expectation values $\langle A_3 B_1\rangle$, $\langle A_3 B_2\rangle$ and $\langle A_3\rangle$  as described in the text. The two antipodal apices of the cones, corresponding to ${\bf x^+}$ and ${\bf x^-}$, and the equator (in black color) can be attained by projective measurements performed on qubits. Whereas, any other point on the surface of the cone (represented by red color) can be achieved by applying POVM measurements.}   
\label{cone}
\end{figure}

\emph{POVM versus projective measurements.}
Now we will show that the quantum body with POVM allowed may contain points that can not be achieved with only projective
measurements. Let us consider the quantum body with $m_A=3$ and $m_B=2$, restricting ourselves to quantum systems of two
dimensional component Hilbert spaces. Let $\hat A_1$, $\hat A_2$, $\hat B_1$ and $\hat B_2$ be the operators, and
$|\psi\rangle$ be the pure maximally entangled state giving the maximum violation of the CHSH inequality. Let the components
$\langle A_i\rangle$, $\langle B_j\rangle$ and $\langle A_iB_j\rangle$ ($i,j=1,2$) of $\bf x$ be derived as the expectation
values of the operators above. Let the components $\langle A_3\rangle$, $\langle A_3B_1\rangle$ and $\langle A_3B_2\rangle$
be the expectation values with $\hat A_3$ corresponding to a projective measurement, that is an observable with eigenvalues
$\pm 1$. Then it can be shown that the region allowed for these three components are the two antipodal apices of the cones
${\bf x}^+$ and ${\bf x}^-$ (when $\hat A_3=\hat I_A$ and $\hat A_3=-\hat I_A$, respectively) and the equator of unit radius (when $\hat A_3$
has both eigenvalues $+1$ and $-1$), shown in Fig.~\ref{cone}. To see this, one has to to use the facts that to get
the other, fixed components of the vector the state must be maximally entangled and the relationship between $\hat B_1$ and
$\hat B_2$ is also well defined. For example, ${\bf x}^\lambda=\lambda {\bf x}^++(1-\lambda){\bf x}^-$ ($0<\lambda<1$), a
point between the antipodes, can not be achieved, as the expectation value of $\hat A_3$ calculated with a
maximally entangled state can only be $\pm 1$ or $0$, and when it is $0$, $\langle A_3B_1\rangle$ and $\langle A_3B_2\rangle$
can not be $0$ at the same time. However, we do achieve the point required with the choice of
$\hat A_3^\lambda=(2\lambda-1)\hat I_A$. This operator corresponds to a POVM with POVM elements $\lambda\hat I_A$ and
$(1-\lambda)\hat I_A$ associated with the +1 and -1 outcome of the measurement, respectively. Similarly, it is easy to prove
that all other points within the red region shown in Fig.~\ref{cone} can be attained with POVM.

\emph{Conclusion.}
We proved that the full set of quantum probabilities in the bipartite scenario generated either by two-outcome projective or by two-outcome POVM
measurements for any given dimension $d$ is concave. However, one may further ask, whether this fact also holds true for more parties and for
more than two outcomes. We also proved that the set generated by projective measurements may be smaller than the one corresponding to
the more general POVM measurements. In case of two-outcome measurements the maximum violation of a Bell inequality with fixed
dimensional systems can still be achieved with projective measurements~\cite{CHTW04,LD}. It remains an open question if this is
true in cases of more than two outcomes~\cite{Gisin07}.

A further question raised by Brunner et al.~\cite{BGS} is that what happens, if we restrict ourselves to measurements on a
given quantum state and look for the set of quantum probabilities generated this way. When we limit the dimensionality
of the Hilbert space, we have shown here that the quantum set is concave, by showing that a point on the convex hull of
points corresponding to deterministic strategies does not belong to the set if the number of measurement settings is
large enough. By restricting ourselves to a particular state, the set can only get smaller, while it will still contain
the points of deterministic strategies.

Finally, it would also be interesting to find out the minimum number of settings which generates a quantum body with a concave shape for a
fixed dimension. In particular, would it be possible in the bipartite case to go below four two-outcome measurement settings per party by local dimension two in order to prove concavity of the corresponding quantum body? 

\acknowledgments
T.~V. has been supported by a J\'anos Bolyai Programme of the Hungarian Academy of Sciences.

\subsection{Appendix: Quantum maximum via SDP}\label{SDP}

We follow the SDP method put forward by Wehner \cite{Wehner} recently, in order to prove analytically quantum bounds for the correlation type Bell inequalities of Eq.~(\ref{bellexpr}). Let us consider the $m\times m$ matrix $M$ with real coefficients 
\begin{equation}
M_{ij}=1-\frac{m}{2}\delta_{ij},
\label{def}
\end{equation} 
introduced in Eq.~(\ref{bellexpr}). As stated in the main text, the expectation values in the polynomial Eq.~(\ref{bellexpr}) can be replaced by dot product of unit vectors, 
\begin{equation}
\max{\sum_{i,j=1}^m{M_{ij}\vec a_i\cdot\vec b_j}},   
\label{q}
\end{equation}
where maximization is taken over all unit vectors $\{\vec a_1,\ldots,\vec a_m,\vec b_1,\ldots,\vec b_m\} \in R^{2m}$. As shown by Tsirelson, the maximum obtained in this way corresponds to the maximum quantum value as well \cite{Tsirelson}.

However, the above problem can be formulated as the following SDP optimization \cite{Wehner}:
\begin{equation}\begin{aligned}\label{primal}
\text{maximize}&\quad \frac{1}{2}\tr{(\Gamma W)}\\
\text{subject to}&\quad
\Gamma\succeq 0,\quad \forall i \,\Gamma_{ii}=1\,.
\end{aligned}\end{equation} 
Here the matrix $W$ is built up as 
\begin{eqnarray}
W = \left(\begin{array}{cc} 
            0 & M\\
            M & 0 
          \end{array}
     \right),
     \label{W}
\end{eqnarray}
and $\Gamma=(\Gamma_{ij})$ is the Gram matrix of the unit vectors $\{\vec a_1,\ldots,\vec a_m,\vec b_1,\ldots,\vec b_m\} \in R^{2m}$. Denoting the columns of the above vectors by $V$, we can write $\Gamma=V^t V$ if and only if $\Gamma$ is positive semidefinite. The constraint $\Gamma_{ii}=1$, on the other hand, owes to the unit length of vectors $\vec a_i$ and $\vec b_j$. Note, that the primal problem defined by (\ref{primal}) is the first step of the hierarchy of semidefinite programs given by Navascu\'{e}s et al. \cite{NPA07,NPA08}.

However, one can also define a dual formulation of the SDP problem (for an exhaustive review see \cite{VB04}):
\begin{equation}\begin{aligned}\label{dual}
\text{maximize}&\quad \tr{(\diag(\lambda))}\\
\text{subject to}&\quad -\frac{1}{2}W + \diag(\lambda)\succeq 0,
\end{aligned}\end{equation} 
where $\lambda$ is a $2m$-dimensional vector with real entries and we note that this dual problem is just the first step of the hierarchy introduced by Doherty et al. \cite{DLTW}.

Let us denote by $p^*$ and $d^*$ the optimal values for the primal and the dual problems, respectively. However, according to weak duality, $d^*\ge p^*$ \cite{VB04}. Thus, in order to prove optimality of the quantum bound one suffices to exhibit a feasible solution both for the primal (\ref{primal}) and for the dual (\ref{dual}) problem and then show that they are in fact equal to each other. For this sake let us guess the primal optimum by setting $\vec a_i, \vec b_j =(1,0,\ldots,0)$ in (\ref{q}) with a Bell matrix defined by (\ref{def}). These vectors correspond to a classical deterministic strategy and this solution yields $p^*=\sum_{i,j}^m{M_{i,j}}=m^2/2$.

Similarly, we guess the solution $\lambda^*=(m/4)(1,\ldots,1)$ for the dual problem, for which the dual value is $d^*=\tr{(\diag(\lambda^*))}=m^2/2$. In order to get a feasible solution, it remains to check according to (\ref{dual}) whether $R=-(1/2)W + \diag(\lambda^*)\succeq 0$ is satisfied. This amounts to prove $\gamma_{min}[R]\ge 0$, where we use the notation $\gamma_{min}[R]$ ($\gamma_{max}[R]$) for the smallest (largest) eigenvalue of a matrix $R$. However, due to Weyl's theorem \cite{HJ85}, for two Hermitian matrices $P$ and $Q$, it holds $\gamma_{min}[P+Q]\ge \gamma_{min}[P]+\gamma_{min}[Q]$. For our particular case, 
\begin{align}
\gamma_{min}[R]&\ge \gamma_{min}[-\frac{1}{2}W]+\gamma_{min}[\diag(\lambda^*)]\nonumber\\&=-\frac{1}{2}\gamma_{max}[W]+\frac{m}{4}.
\label{gamma} 
\end{align}  
The eigenvalues of matrix $W$ in the form (\ref{W}) are given by the singular values $\sigma_s=\sqrt{\gamma_s \gamma_s^*}$ of matrix $M$ of (\ref{def}) and their negatives. The eigenvalues  of $M$ on the other hand are the roots of the characteristic polynomial $\mathrm{det}(M-\gamma_s\one)$, where $\one$  is the 
$m\times m$ unit matrix. In \cite{VP09} we found that the determinant of an $m\times m$ matrix with diagonal elements $p$ and non-diagonal elements $q$ is
$[p+(m-1)q](p-q)^{m-1}$. By inserting $p=1-m/2-\gamma_s$ and $q=1$ into the determinant above, we obtain the roots $\gamma_s=\pm m/2$. 
This result implies $\gamma_{max}[W]=\frac{m}{2}$. By substituting this value into (\ref{gamma}) we get $\gamma_{min}[R]\ge 0$. This implies that this solution for $d^*$ is feasible, and recalling the guessed solution $p^*$, we have $d^*=p^*$. Thus the maximum quantum value of the Bell polynomial $M$ defined by Eq.~(\ref{bellexpr}) is equal to $m^2/2$, which can be achieved by classical means as well.

\end{document}